\documentclass[aps,prc,reprint,superscriptaddress]{revtex4-1}

\usepackage{graphicx,topcapt,booktabs,SIunits,commath,units,mhchem,amsmath}

\begin{document}

\title{New $\gamma$-ray Transitions Observed in $^{19}$Ne with Implications for the $^{15}$O($\alpha$,$\gamma$)$^{19}$Ne Reaction Rate}

\author{M.R.~Hall}
\email{mhall12@alumni.nd.edu}
\author{D.W.~Bardayan}
\affiliation{Department of Physics, University of Notre Dame, Notre Dame, Indiana 46556, USA}
\author{T.~Baugher}
\affiliation{Department of Physics and Astronomy, Rutgers University, New Brunswick, New Jersey 08903, USA}
\author{A.~Lepailleur}
\affiliation{Department of Physics and Astronomy, Rutgers University, New Brunswick, New Jersey 08903, USA}
\author{S.D.~Pain}
\affiliation{Physics Division, Oak Ridge National Laboratory, Oak Ridge, Tennessee 37831, USA}
\author{A.~Ratkiewicz}
\affiliation{Department of Physics and Astronomy, Rutgers University, New Brunswick, New Jersey 08903, USA}

\author{S.~Ahn}
\affiliation{National Superconducting Cyclotron Laboratory, Michigan State University, East Lansing, Michigan 48824, USA}

\author{J.M.~Allen}
\affiliation{Department of Physics, University of Notre Dame, Notre Dame, Indiana 46556, USA}

\author{J.T.~Anderson}
\affiliation{Physics Division, Argonne National Laboratory, Argonne, Illinois 60439, USA}

\author{A.D.~Ayangeakaa}
\affiliation{Physics Division, Argonne National Laboratory, Argonne, Illinois 60439, USA}

\author{J.C.~Blackmon}
\affiliation{Department of Physics and Astronomy, Louisiana State University, Baton Rouge, Louisiana 70803, USA}

\author{S.~Burcher}
\affiliation{Department of Physics and Astronomy, University of Tennessee, Knoxville, Tennessee 37996, USA}

\author{M.P.~Carpenter}
\affiliation{Physics Division, Argonne National Laboratory, Argonne, Illinois 60439, USA}

\author{S.M.~Cha}
\affiliation{Department of Physics, Sungkyunkwan University, Suwon 16419, South Korea}

\author{K.Y.~Chae}
\affiliation{Department of Physics, Sungkyunkwan University, Suwon 16419, South Korea}

\author{K.A.~Chipps}
\affiliation{Physics Division, Oak Ridge National Laboratory, Oak Ridge, Tennessee 37831, USA}

\author{J.A.~Cizewski}
\affiliation{Department of Physics and Astronomy, Rutgers University, New Brunswick, New Jersey 08903, USA}

\author{M.~Febbraro}
\affiliation{Physics Division, Oak Ridge National Laboratory, Oak Ridge, Tennessee 37831, USA}

\author{O.~Hall}
\affiliation{Department of Physics, University of Notre Dame, Notre Dame, Indiana 46556, USA}
\affiliation{Department of Physics, University of Surrey, Guildford, Surrey GU2 7XH, United Kingdom}

\author{J.~Hu}
\affiliation{Department of Physics, University of Notre Dame, Notre Dame, Indiana 46556, USA}

\author{C.L.~Jiang}
\affiliation{Physics Division, Argonne National Laboratory, Argonne, Illinois 60439, USA}

\author{K.L.~Jones}
\affiliation{Department of Physics and Astronomy, University of Tennessee, Knoxville, Tennessee 37996, USA}

\author{E.J.~Lee}
\affiliation{Department of Physics, Sungkyunkwan University, Suwon 16419, South Korea}

\author{P.D.~O'Malley}
\affiliation{Department of Physics, University of Notre Dame, Notre Dame, Indiana 46556, USA}

\author{S.~Ota}
\affiliation{Physics Division, Lawrence Livermore National Laboratory, Livermore, California 94551, USA}

\author{B.C.~Rasco}
\affiliation{Department of Physics and Astronomy, Louisiana State University, Baton Rouge, Louisiana 70803, USA}

\author{D.~Santiago-Gonzalez}
\affiliation{Department of Physics and Astronomy, Louisiana State University, Baton Rouge, Louisiana 70803, USA}

\author{D.~Seweryniak}
\affiliation{Physics Division, Argonne National Laboratory, Argonne, Illinois 60439, USA}

\author{H.~Sims}
\affiliation{Department of Physics and Astronomy, Rutgers University, New Brunswick, New Jersey 08903, USA}
\affiliation{Department of Physics, University of Surrey, Guildford, Surrey GU2 7XH, United Kingdom}

\author{K.~Smith}
\affiliation{Department of Physics and Astronomy, University of Tennessee, Knoxville, Tennessee 37996, USA}

\author{W.P.~Tan}
\affiliation{Department of Physics, University of Notre Dame, Notre Dame, Indiana 46556, USA}

\author{P.~Thompson}
\affiliation{Physics Division, Oak Ridge National Laboratory, Oak Ridge, Tennessee 37831, USA}
\affiliation{Department of Physics and Astronomy, University of Tennessee, Knoxville, Tennessee 37996, USA}

\author{C.~Thornsberry}
\affiliation{Department of Physics and Astronomy, University of Tennessee, Knoxville, Tennessee 37996, USA}

\author{R.L.~Varner}
\affiliation{Physics Division, Oak Ridge National Laboratory, Oak Ridge, Tennessee 37831, USA}

\author{D.~Walter}
\affiliation{Department of Physics and Astronomy, Rutgers University, New Brunswick, New Jersey 08903, USA}

\author{G.L.~Wilson}
\affiliation{Department of Physics and Astronomy, Louisiana State University, Baton Rouge, Louisiana 70803, USA}
\affiliation{Department of Physics and Applied Physics, University of Massachusetts Lowell, Lowell, Massachusetts 01854, USA}

\author{S.~Zhu}
\affiliation{Physics Division, Argonne National Laboratory, Argonne, Illinois 60439, USA}

\date{\today}

\begin{abstract}
The $^{15}$O($\alpha$,$\gamma$)$^{19}$Ne reaction is responsible for breakout from the hot CNO cycle in Type I x-ray bursts. Understanding the properties of resonances between $E_x = 4$ and 5 MeV in $^{19}$Ne is crucial in the calculation of this reaction rate. The spins and parities of these states are well known, with the exception of the 4.14- and 4.20-MeV states, which have adopted spin-parities of 9/2$^-$ and 7/2$^-$, respectively. Gamma-ray transitions from these states were studied using triton-$\gamma$-$\gamma$ coincidences from the $^{19}$F($^{3}$He,$t\gamma$)$^{19}$Ne reaction measured with GODDESS (Gammasphere ORRUBA Dual Detectors for Experimental Structure Studies) at Argonne National Laboratory. The observed transitions from the 4.14- and 4.20-MeV states provide strong evidence that the $J^\pi$ values are actually 7/2$^-$ and 9/2$^-$, respectively. These assignments are consistent with the values in the $^{19}$F mirror nucleus and in contrast to previously accepted assignments. 

\end{abstract}

\maketitle{}

\section{Introduction}
The $^{15}$O($\alpha$,$\gamma$)$^{19}$Ne reaction is an important breakout reaction from the hot CNO cycle in explosive astrophysical environments such as Type I x-ray bursts (XRBs). Type I XRBs are thought to occur in close binary systems containing an accreting neutron star \cite{Woosley1976, Joss1977}. The hydrogen-rich material accreted onto the surface of the star provides the fuel for the hot CNO cycle, which can then break out into the \textit{rp}-process, synthesizing isotopes up to $A\approx100$ \cite{Schatz2001}. Knowledge of the $^{15}$O($\alpha$,$\gamma$)$^{19}$Ne breakout reaction is therefore critical to our understanding of the nucleosynthesis occurring in this environment. It has been shown that this reaction rate has a large effect on the light curves observed from XRBs, and models do not even predict explosions if the rate is near the lower limit of its uncertainty \cite{Fisker2006,Cyburt2016}. The reaction can not be measured directly in the important astrophysical temperature range due to the currently insufficient intensity of radioactive $^{15}$O beams and small reaction cross section. Therefore, the rate must be estimated by measuring the properties of the important resonances in $^{19}$Ne.

The resonances in the reaction cross section correspond to the energy levels in $^{19}$Ne above the alpha separation threshold at $S_\alpha = 3.529$ MeV. Many of the resonances from energy levels between 4 and 5 MeV have been characterized in previous experiments \cite{Mythili2008,Tan2009,Davids2011,Parikh2015}. However, the spins of two $^{19}$Ne states at 4.14 and 4.20 MeV remain in question. 

Over 45 years ago \cite{Garrett1972}, these two levels were proposed as members of the $K^\pi=1/2^-$ rotational band, with negative parity, and the mirrors of the 3.998- and 4.032-MeV levels, which have $J^\pi$ values of 7/2$^-$ and 9/2$^-$, respectively \cite{Tilley1995}. A study of the $^{16}$O($^6$Li,$t$)$^{19}$Ne reaction by Garrett \textit{et al.} \cite{Garrett1972} first showed that the 4.14- and 4.20-MeV $^{19}$Ne states had spin-parities of 7/2$^-$ or 9/2$^-$ and suggested that their assignments were reversed from their order in the $^{19}$F mirror nucleus. Since that time, evidence supporting both spin-parity assignments for the 4.14- and 4.20-MeV states has been found; the adopted \cite{Tilley1995} spin assignments remain uncertain.

\begin{figure}[ht!]
\begin{center}
\includegraphics[width=\linewidth]{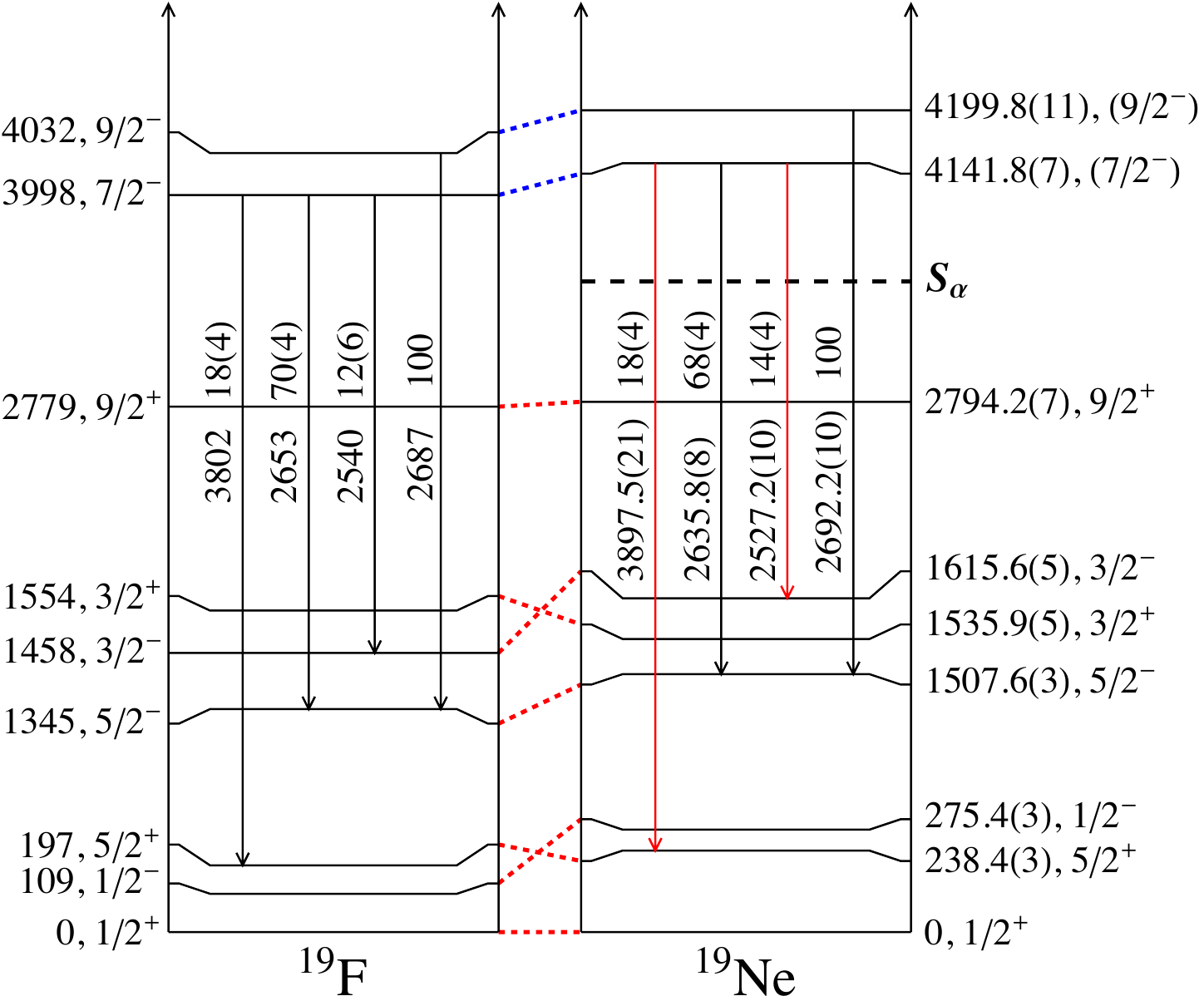}

\caption{(Color online) Partial level schemes of $^{19}$F and $^{19}$Ne highlighting two states near 4 MeV and showing mirror connections between levels (dashed lines); the mirror connections updated in this work are shown in blue. The $^{19}$F transitions, branching ratios (\%), and energies (keV) are from Ref. \cite{Tilley1995}. The $^{19}$Ne 4.14- and 4.20-MeV state $\gamma$-ray transitions, branching ratios, and energies were determined in this work. The two red transitions were first observed in this measurement.}

\label{Lev}
\end{center}
\end{figure}

Gamma rays from the decay of the $^{19}$Ne 4.14- and 4.20-MeV states were studied by Davidson \textit{et al.} \cite{Davidson1973} using the $^{17}$O($^3$He,$n\gamma$)$^{19}$Ne reaction, where they reported the observation of three transitions. For the 4.14-MeV state, a single transition to the 1.508-MeV state was observed, whereas for the 4.20-MeV state, transitions to the 0.238- and 1.508-MeV states were reported. Based on the (relatively weak) transition to the 0.238-MeV 5/2$^+$ state, the 4.20-MeV state was assigned $J^\pi=7/2^-$, consistent with the $J^\pi$ assignment of the 3.998-MeV state and transition to the 0.197-MeV state in the $^{19}$F mirror nucleus (see Fig. \ref{Lev}). The analysis of triton angular distributions from the $^{19}$F($^3$He,$t$)$^{19}$Ne reaction study by Parikh \textit{et al.} \cite{Parikh2015} was also consistent with multi-step \textsc{FRESCO} calculations for a 9/2$^-$ assignment for the 4.14-MeV state and a 7/2$^-$ assignment for the 4.20-MeV state.

However, Some evidence suggests that the spin assignments for the 4.14- and 4.20-MeV states could be reversed and, therefore, in the same order that they occur in $^{19}$F. The lifetimes and $\alpha$-decay branching ratios of the states in $^{19}$Ne that are important in the $^{15}$O($\alpha$,$\gamma$)$^{19}$Ne reaction were measured at the University of Notre Dame using the $^{17}$O($^3$He,$n$)$^{19}$Ne and $^{19}$F($^3$He,$t$)$^{19}$Ne reactions, respectively \cite{Tan2005,Tan2007,Tan2009}. For the 4.14- and 4.20-MeV states, the lifetimes were measured to be 18$^{+2}_{-3}$ fs and 43$^{+12}_{-9}$ fs, respectively \cite{Tan2009, Tan2005}. A comparison with the measured lifetimes of the 3.998- ($\tau=19\pm$7 fs) and 4.032-MeV ($\tau=67\pm$15 fs) $^{19}$F states suggested that the spin-parities of the 4.14- and 4.20-MeV states in $^{19}$Ne should be 7/2$^-$ and 9/2$^-$, respectively, analogous to the $^{19}$F mirror nucleus. It was also noted by Ref. \cite{Tan2009} that the resonance corresponding to the 4.14-MeV state may dominate the $^{15}$O($\alpha$,$\gamma$)$^{19}$Ne reaction rate in a narrow temperature range around 0.8 GK if the state has a sufficient $\alpha$-decay branching ratio.

In addition, in a study of the $^{15}$O($\alpha$,$\gamma$)$^{19}$Ne reaction rate by Davids \textit{et al.} \cite{Davids2011}, the reduced transition probabilities of the 4.14- and 4.20-MeV levels were calculated and compared with those found for the 3.998-MeV state in $^{19}$F. A transition to the 1.508-MeV ($J^\pi = 5/2^-$) state, to which both $^{19}$Ne levels primarily decay, will be either an $M$1 or $E$2 transition depending on the spin-parity. For the $^{19}$F states at 3.998- and 4.032-MeV, the $B(M1)$ and $B(E2)$ values are 0.0017$^{+0.0010}_{-0.0005}$ MeV fm$^3$ and 90$\pm$20 MeV fm$^5$, respectively. If the spin-parity of the 4.14-MeV state is assumed to be 7/2$^-$, this yields $B(M1) = 0.0024^{+0.0010}_{-0.0009}$ MeV fm$^3$, which is in good agreement with the $^{19}$F value. Similarly, if the 4.20-MeV state is assumed to be 9/2$^-$ yields $B(E2) = 150^{+60}_{-50}$ MeV fm$^5$, which is also in good agreement with $^{19}$F. The authors note that the reduced transition probabilities calculated with opposite spin assignments did not agree, but the measured $\gamma$-ray branching ratios still supported the  tentative spin assignments adopted in Ref. \cite{Tilley1995}. 

\section{Experimental Setup and Analysis}
\begin{figure}[ht!]
\begin{center}
\includegraphics[width=.9\linewidth]{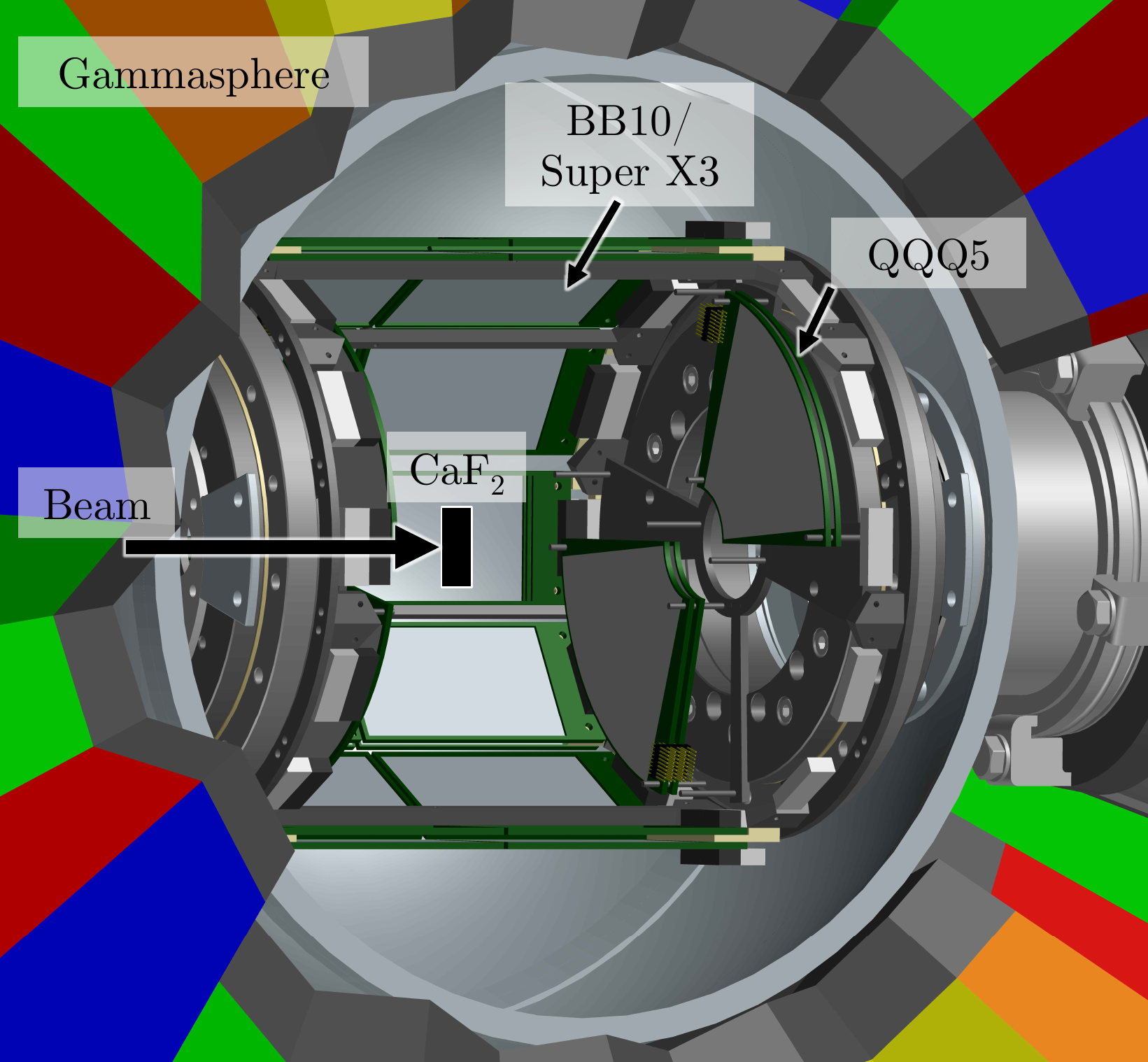}

\caption{(Color online) Rendering of the GODDESS setup showing the beam direction, target location, and ORRUBA in position inside Gammasphere \cite{Rat2013}.}

\label{Setup}
\end{center}
\end{figure}

To resolve these discrepancies of the $J^\pi$ assignments of the 4.14- and 4.20-MeV levels in $^{19}$Ne, the $^{19}$F($^3$He,$t\gamma$)$^{19}$Ne reaction was measured at Argonne National Laboratory using the coupling of the Compton-suppressed high-purity germanium (HPGe) detector array Gammasphere \cite{Lee1992} with the silicon detector array ORRUBA (Oak Ridge Rutgers University Barrel Array) \cite{Pain2007}, called Gammasphere ORRUBA Dual Detectors for Experimental Structure Studies (GODDESS) \cite{Rat2013,Pain2014,Pain2017}. A 30-MeV $^{3}$He beam was delivered by the ATLAS accelerator onto a 938-$\mu$g/cm$^2$ CaF$_2$ target at the GODDESS target position. A rendering \cite{Rat2013} of the GODDESS setup can be seen in Fig. \ref{Setup}; a more in-depth description can be found in Ref. \cite{Hall2019}. 

\begin{figure}[ht!]
\begin{center}
\includegraphics[width=\linewidth]{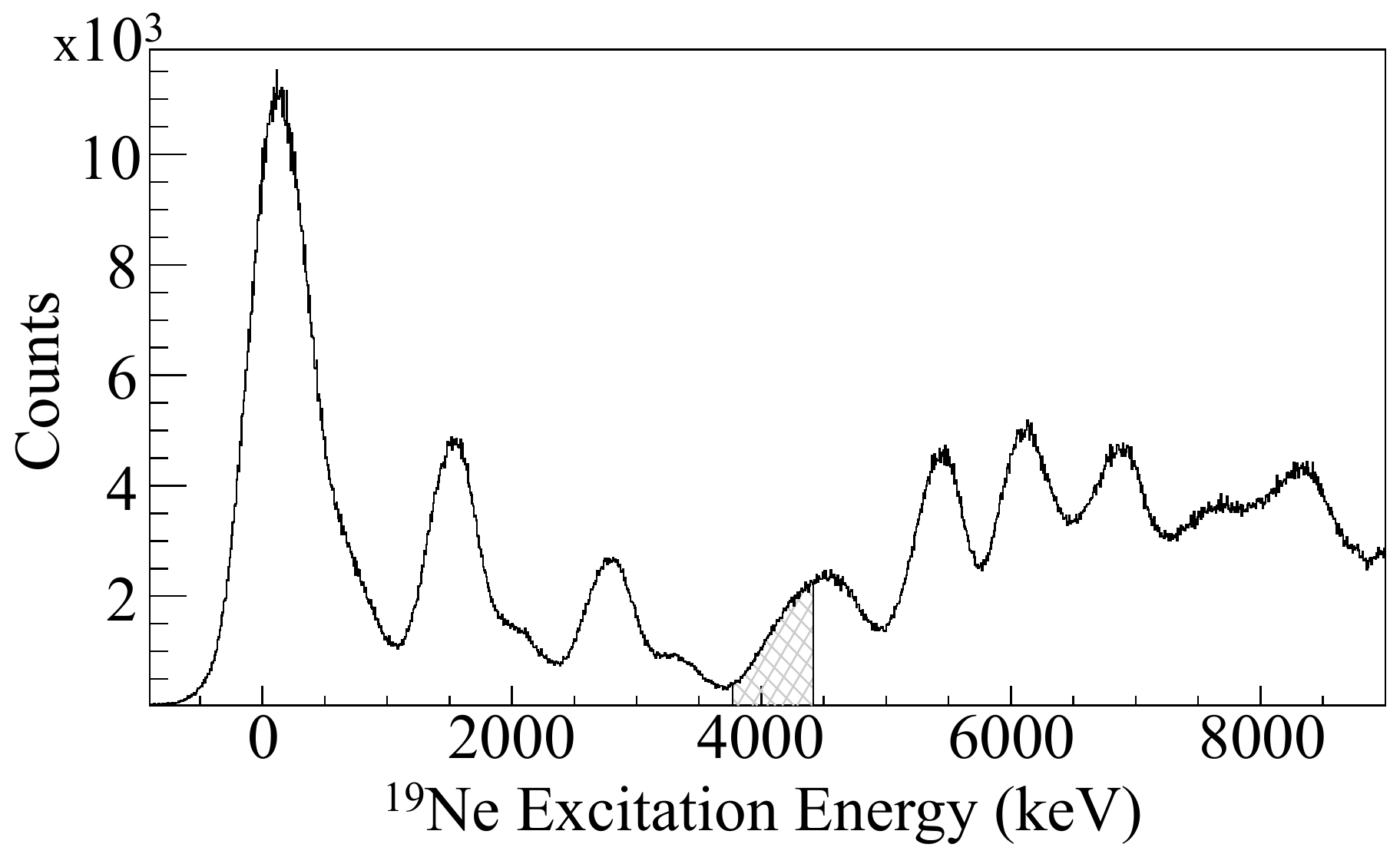}

\caption{Triton spectrum populating excitations in $^{19}$Ne at $\theta_{lab}=20^\circ$. The shaded region corresponds to $^{19}$Ne excitation energies between 3.8 and 4.4 MeV, which were gated on to produce the results shown.}

\label{Exspec}
\end{center}
\end{figure}

The charged particles produced in the reaction were detected in $\Delta$E-E telescopes in the downstream half of ORRUBA. In the barrel, the six telescopes consisted of a 65-$\mu$m-thick BB10 detector in front of a 1000-$\mu$m-thick Super X3 detector. Downstream of the ORRUBA barrel, an endcap of two QQQ5 telescopes, consisting of highly-segmented detectors with thicknesses of 100 and 1000 $\mu$m, was mounted. A 0.5-mm-thick aluminum plate was mounted in front of the QQQ5 detectors to stop the elastically-scattered $^3$He beam. On average, the plate reduced the triton energies in the endcap detectors by approximately 1/3, allowing the tritons from the population of the $^{19}$Ne ground state to stop in the QQQ5 telescopes. In total, ORRUBA covered laboratory angles ranging from approximately 18$^\circ$ to 162$^\circ$ (though, only laboratory angles less than 90$^\circ$ were considered during the analysis). The triton spectrum populating excitations in $^{19}$Ne can be seen in Fig. \ref{Exspec}. This spectrum looks different than that of Ref. \cite{Tan2007} due to the different bombarding energy, different angular coverage of the detectors, and the existence of the previously mentioned aluminum plate in front of the detectors. 

Gamma rays from the decay of $^{19}$Ne were measured in Gammasphere, in coincidence with the tritons from the reaction between $^{19}$Ne excitation energies of 3.8 and 4.4 MeV (shaded region in Fig. \ref{Exspec}). To calibrate Gammasphere, sources of $^{152}$Eu, $^{56}$Co, and $^{238}$Pu+$^{13}$C were used, which provided calibration $\gamma$ rays with energies ranging from 122 to 6128~keV. The systematic uncertainties on the energy calibration were estimated to range from $0.3$ to $2.0$~keV between energies of 100 and 7000 keV. These uncertainties were combined in quadrature with the statistical uncertainties on each of the peak centroids to determine the uncertainty on the transition energy. The $^{19}$Ne excitation energies were calculated for each detected $\gamma$-ray cascade if more than one cascade was placed. The final excitation energy was determined by averaging each value and weighting them by their uncertainties.

Since the lifetimes of the 4.14- and 4.20-MeV levels are very short, the decay of these states occurred when the $^{19}$Ne nuclei were in flight. Therefore, the $\gamma$ rays from the de-excitation of these states were found to be Doppler broadened and a Doppler correction was applied to the Gammasphere spectra for the transitions depopulating the 4.14- and 4.20-MeV states. The angle and energy of the $^{19}$Ne nuclei were calculated for each event using the angle and energy of the triton detected in ORRUBA, and the values of $\beta$ used for the correction ranged between 0.005 and 0.025. The correction was applied assuming the $^{19}$Ne nuclei did not lose any energy in the target before decaying, since this assumption produced $\gamma$-ray peaks with the best energy resolution and signal-to-noise ratio. 

To further improve the signal-to-noise ratio, the differences between the recorded Gammasphere and ORRUBA time stamps for each event were used to reduce the random $\gamma$-ray background. True coincidences appear as a sharp peak in this time difference spectrum. Off-peak timing was used to estimate the random-coincident background present in the spectra. The random background generated using the timing was subtracted from the spectra to produce the results presented in the following section. 

\section{Results}

In total, four transitions from the $^{19}$Ne states at 4.14 and 4.20 MeV were identified in the data via triton-$\gamma$-$\gamma$ coincidences. Using the transition energies, the levels were determined to have energies of 4141.8$\pm$0.7 keV and 4199.8$\pm$1.1 keV, which are in good agreement with previous measurements \cite{Tilley1995}. A comparison between the $^{19}$F and $^{19}$Ne partial level schemes and observed transitions is displayed in Fig. \ref{Lev}. 

\begin{figure}[ht!]
\begin{center}
\includegraphics[width=.8\linewidth]{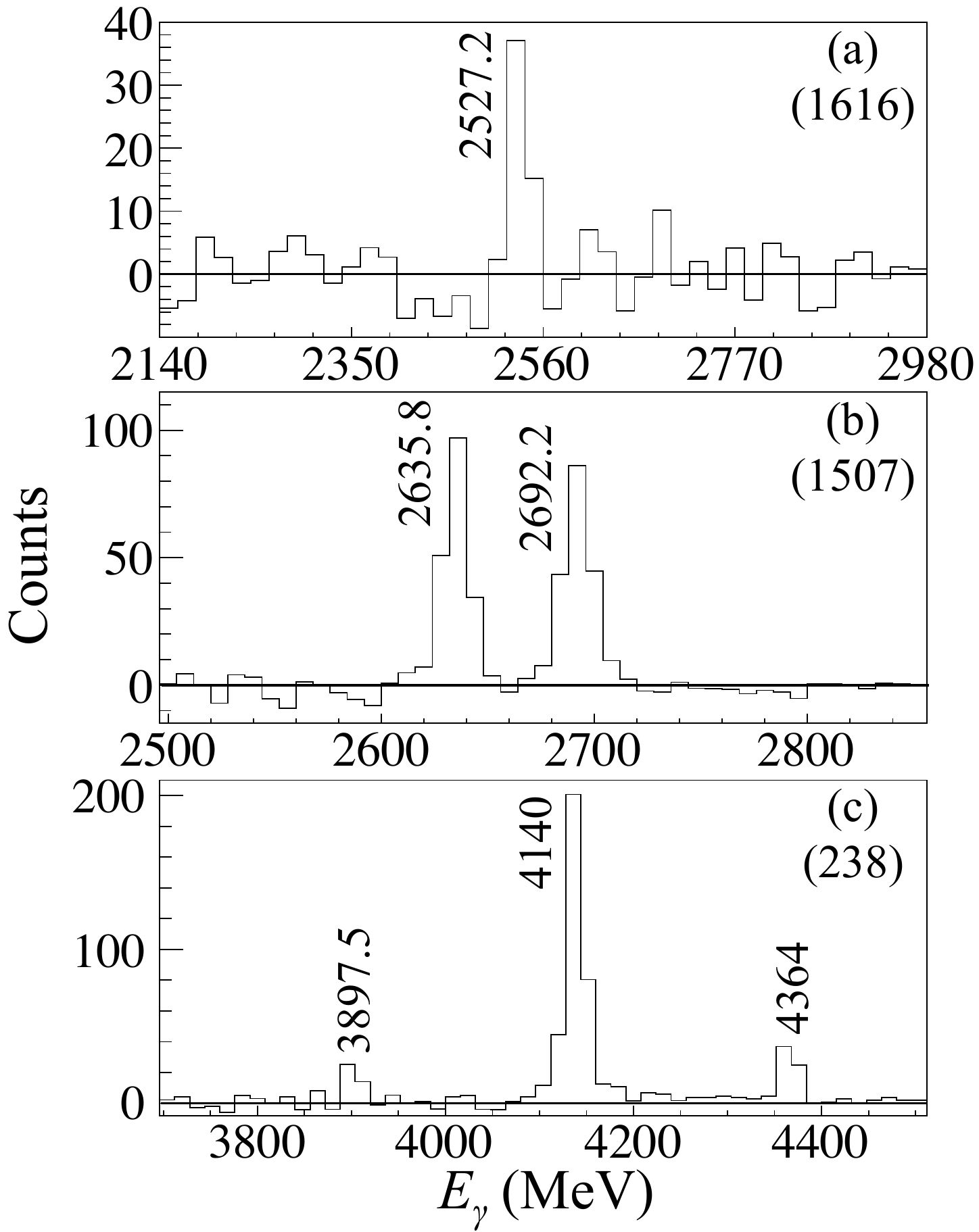}

\caption{Random-subtracted Gammasphere spectra generated by gating on tritons corresponding to $^{19}$Ne excitation energies between 3.8 and 4.4 MeV and $\gamma$-$\gamma$ coincidences. (a) Gated on the 1340-, 1377- and 1616-keV transitions depopulating the 1616-keV 3/2- state. The 2527.2-keV transition from the 4141.8-keV state is observed for the first time. (b) Gated on the 1232- and 1269-keV transitions depopulating the 1507-keV 5/2$^-$ state. (c) Gated on the 238-keV 5/2$^{+}$ to ground-state transition. The 3897.5-keV transition from the 4141.8-keV state is observed for the first time. The transitions labeled 4140 and 4364 keV are previously-observed transitions from the 4378- and 4602-keV levels. The binning of the histograms is 20 keV/bin, 8 keV/bin, and 16 keV/bin, respectively.}

\label{data}
\end{center}
\end{figure}

Figure \ref{data} summarizes the justification for the placement of the transitions depopulating the 4141.8- and 4199.8-keV states. For the 4141.8-keV state, three transitions were observed in the triton-gated $\gamma$-ray spectra. Figure 3a shows the 2527.2(10)-keV $\gamma$ ray from the de-excitation of the 4141.8-keV state, which was produced by gating on the $\gamma$ rays from the de-excitation of the 1616-keV state. Figure 3b is gated on the two transitions that depopulate the 5/2$^-$ 1507-keV state; the two transitions observed depopulate the 4141.8- and 4199.8-keV levels. The $\gamma$-ray spectrum shown in Figure 3c is gated on the 238-keV 5/2$^+$ to ground state transition, confirming the 3897.5-keV transition depopulating the 4141.8-keV state. The branching ratios for the transitions from the 4141.8-keV state were determined to be 14(4)\%, 68(4)\%, and 18(4)\%, respectively.

In contrast to Ref. \cite{Davidson1973}, there is no evidence in the Fig. 3c spectrum of a 3962-keV de-excitation from the 4199.8-keV state to the 238-keV state. Since this spectrum was also gated on tritons corresponding to excitation energies between 3.8 and 4.4 MeV, if this transition did exist it should have been visible in this spectrum. In Ref. \cite{Davidson1973}, this transition is relatively weak in a spectrum only gated on neutrons, with no excitation energy gate and no $\gamma$-$\gamma$ coincidences. Therefore, it is likely that the previously observed, weak transition was incorrectly placed as depopulating the 4199.8-keV state.

The de-excitations from the 4141.8-keV state to the 238- and 1616-keV states were first observed in this work. The low spin-parity of the 1616-keV state ($J^\pi = 3/2^-$) suggests that the spin-parity of the 4141.8-keV state is 7/2$^-$, instead of 9/2$^-$, based on the multipolarity of the transition. In addition, an $E2$ (rather than higher multipolarity) transition is consistent with the measured (short) lifetime of this state. Since one 7/2$^-$ and one 9/2$^-$ state is expected in this region, the spin-parity of the 4199.8-keV state must be 9/2$^-$. These spin-parity assignments are also supported by the transitions previously observed for the mirror states in $^{19}$F (see Fig. \ref{Lev}). The $\gamma$-ray branching ratios obtained for the 4141.8-keV state also match well with those found for the $J^\pi=7/2^-$ 3998-keV state \cite{Tilley1995}: 18(4)\%, 70(4)\%, and 12(6)\% for the 3998-keV $^{19}$F state compared to 18(4)\%, 68(4)\%, and 14(4)\% for the 4141.8-keV $^{19}$Ne state. 

\begin{figure}[ht!]
\begin{center}
\includegraphics[width=\linewidth]{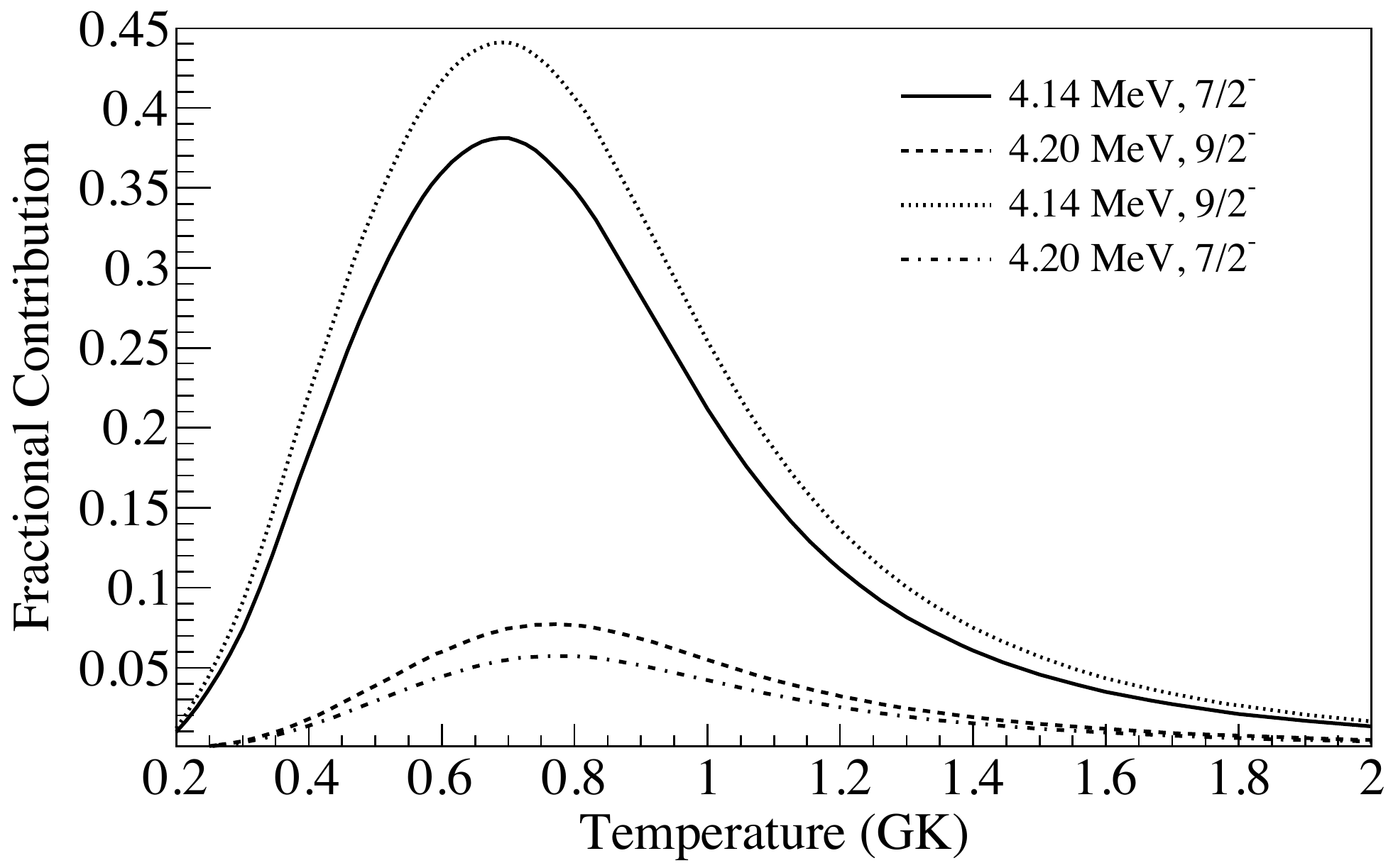}

\caption{Fractional contributions to the $^{15}$O($\alpha$,$\gamma$)$^{19}$Ne reaction rate for the 4.14 and 4.20-MeV states assuming the two sets of spin-parity assignments and  branching ratios of $B(\alpha) = 1.2\times10^{-3}$ \cite{Tan2009}.}

\label{rate}
\end{center}
\end{figure}

Figure \ref{rate} shows the fractional contributions of the 4.14- and 4.20-MeV states to the $^{15}$O($\alpha$,$\gamma$)$^{19}$Ne reaction rate. The calculated fractional contributions assume $\alpha$-decay branching ratios of $B(\alpha) = 1.2\times10^{-3}$, as found in Ref. \cite{Tan2009}. From Fig. \ref{rate}, it is clear that the 4.14-MeV state has less importance than considered previously with a spin of 7/2$^-$, while the 4.20-MeV state is slightly more important.

\section{Conclusion}

The $^{19}$F($^3$He,$t\gamma$)$^{19}$Ne reaction was measured with GODDESS to provide additional information on $^{19}$Ne excitations important in nucleosynthesis. The 4.14- and 4.20-MeV states in $^{19}$Ne could provide important resonances for the $^{15}$O($\alpha$,$\gamma$)$^{19}$Ne breakout reaction in Type I x-ray bursts. However, conflicting information regarding the spin-parities of these states made their potential contributions uncertain. The $^{19}$F($^{3}$He,$t\gamma$)$^{19}$Ne reaction was studied using GODDESS to search for $\gamma$-ray transitions that could resolve this discrepancy.

Using triton-$\gamma$-$\gamma$ coincidences, the two levels were confirmed at energies of 4141.8 and 4199.8 keV. Two new transitions were observed from the 4141.8-keV state to the 238- and 1616-keV states. In addition, two previously observed transitions were also found from the 4141.8- and 4199.8-keV states to the 1508-keV state. The decay scheme from these states matches well with the decay scheme previously observed for the two proposed mirror states in $^{19}$F. The present triton-gated $\gamma$-ray measurements and the results from Refs. \cite{Tan2009,Davids2011} suggest that the previously accepted spin-parities for these states should be reversed. Therefore, we assign spin-parities of 7/2$^-$ and 9/2$^-$ to the 4141.8- and 4199.8-keV states, respectively. It was noted in Ref. \cite{Tan2009} that the 4141.8-keV state could have the largest contribution to the $^{15}$O($\alpha$,$\gamma$)$^{19}$Ne reaction rate if it has a sufficient $\alpha$-decay branching ratio. Further studies targeting this quantity are necessary to help constrain the rate further.

\section{Acknowledgements}

This research was supported in part by the National Science Foundation Grant Numbers PHY-1419765 (Notre Dame) and PHY-1404218 (Rutgers), the National Nuclear Security Administration under the Stewardship Science Academic Alliances program through DOE Cooperative Agreement DE-NA002132, and by the National Research Foundation of Korea (NRF) grant funded by the Korea government (MSIT) (numbers NRF-2016R1A5A1013277 and NRF-2013M7A1A1075764). The authors also acknowledge support from the DOE Office of Science, Office of Nuclear Physics, under contract numbers DE-AC05-00OR22725, DE-FG02-96ER40963, DE-FG02-96ER40978, and DE-AC02-06CH11357. This research used resources of Argonne National Laboratory’s ATLAS facility, which is a DOE Office of Science User Facility.

\bibstyle{apsrev4}

\end{document}